\documentclass[conference]{IEEEtran}
\IEEEoverridecommandlockouts

\usepackage{cite}
\usepackage{amsmath,amssymb,amsfonts}
\usepackage{algorithmic}
\usepackage{graphicx}
\usepackage{textcomp}
\usepackage{xcolor}
\usepackage{soul}
\usepackage{pdfpages}
\usepackage{bbm}
\usepackage{dsfont}
\usepackage{hyperref}
\usepackage{siunitx}
\usepackage{gensymb}
\def\BibTeX{{\rm B\kern-.05em{\sc i\kern-.025em b}\kern-.08em
    T\kern-.1667em\lower.7ex\hbox{E}\kern-.125emX}}
\begin{document}

\title{Visually Supervised Speaker Detection and Localization via Microphone Array\\
\thanks{Work supported by InnovateUK (105168) `Polymersive: Immersive video production tools for studio and live events'. 

978-1-6654-3288-7/21/\$31.00 ©2021 IEEE

\hl{Erratum: Due to a bug in the evaluation script, the correct average distance (aD) metric is here reported in yellow. The analysis remains unchanged from the original paper as the trend between the old and new measures are perfectly monotonic. The bug was caused by an incorrect normalization factor.}}
}

\author{\IEEEauthorblockN{Davide Berghi, Adrian Hilton and Philip J.B. Jackson}
\IEEEauthorblockA{\textit{CVSSP, University of Surrey, UK} \\
\{d.berghi, a.hilton, p.jackson\}@surrey.ac.uk}
}
\maketitle

\begin{abstract}
Active speaker detection (ASD) is a multi-modal task that aims to identify who, if anyone, is speaking from a set of candidates. Current audio-visual approaches for ASD typically rely on visually pre-extracted face tracks (sequences of consecutive face crops) and the respective monaural audio. However, their recall rate is often low as only the visible faces are included in the set of candidates. Monaural audio may successfully detect the presence of speech activity but fails in localizing the speaker due to the lack of spatial cues. Our solution extends the audio front-end using a microphone array. We train an audio convolutional neural network (CNN) in combination with beamforming techniques to regress the speaker’s horizontal position directly in the video frames. 
We propose to generate weak labels using a pre-trained active speaker detector on pre-extracted face tracks. Our pipeline embraces the ``student-teacher'' paradigm, where a trained ``teacher'' network is used to produce pseudo-labels visually. The ``student'' network is an audio network trained to generate the same results. At inference, the student network can independently localize the speaker in the visual frames directly from the audio input. 
Experimental results on newly collected data prove that our approach significantly outperforms a variety of other baselines as well as the teacher network itself. It results in an excellent speech activity detector too.  
\end{abstract}


\section{Introduction}
As the human brain explores the world, it localizes objects in the environment integrating multi-modal signals into a common reference frame. Visual and auditory cues play a key role in formation of a coherent localization.
Thus, we have learned to localize audio sources, with a certain accuracy, thanks to subtle differences in the acoustic signals picked up by our auditory system, and to integrate them into our perceived model of the world. 
Our ability to integrate audio and visual cues as a single percept in a common reference frame improves reliability and robustness, maintaining viable estimates when a modality is corrupted or unavailable.

Audio-visual systems for speaker detection and localization arise in applications such as speaker diarization, surveillance, camera steering and live media production. 
In practice, camera and microphone sensors present different yet complementary features, e.g.: their size, cost, power consumption, field of view, temporal and spatial resolutions, robustness to changes in light and occlusions.
A good audio-visual system must be designed to take full advantage of these complementary features but also to rely on a single modality when its counterpart is missing.
This work investigates whether these systems can benefit from joint audio-visual training to gain performance when the visual input fails.


Annotating the ground truth speaker position, as is required for supervised machine learning, is expensive and time-consuming. In contrast, our semi-supervised approach uses manually-screened, automatic audio-visual `pseudo labels' to supervise the training of an audio-only network to detect and locate an active speaker.
Thus, we seek to leverage unlabeled audio-visual training data to learn how to localize an active speaker in the visual reference frame, purely from multi-channel audio signals. 
The following sections present the background (\ref{bg}), the proposed approach (\ref{appr}), the experiments (\ref{exp}), the results achieved (\ref{res}), and our conclusions (\ref{concl}).

\section{Background} \label{bg}

\subsection{Self-Supervised Audio-Visual Learning}

Self-supervised audio-visual learning is a research area that is rapidly gaining interest. It provides not only multi-modal solutions to tackle traditional problems, e.g., exploiting 
visual information for speech enhancement and separation \cite{Afouras18} or solving the `cocktail party' problem \cite{Ephrat:2018:Looking2Listen}, but
it has also given rise to new tasks: e.g., \cite{morgadoNIPS18} and \cite{Gao:2019:visualsound} 
proposed self-supervised approaches to generate spatial audio from videos with monaural sound. 
Other works exploited the co-occurrence of audio and visual events to localize and separate audio sources that were seen in the video \cite{gao2018objectSounds,Gao:2019:coseparating,zhao:2018:ECCV}.
However, these approaches require both modalities present, which fail with poor lighting or visual occlusion. 
In contrast, our method requires the visual modality only during training, 
drawing on the student-teacher paradigm \cite{Hinton2015DistillingTheKnowledge,Aytar:2016:soundNet,Owens2016AmbientSP}. 
In audio-visual learning, this paradigm exploits the natural synchronization of audio and visual signals as a bridge between modalities, enabling one modality to supervise another \cite{Owens2016AmbientSP,Aytar:2016:soundNet,Arandjelovic:2017:look}.
Closest to our work, \cite{Gan2019SelfSupervisedMV,valverde:2021:mmdistillnet,vasudevan:2020:semanticobject} adopted a student-teacher approach to detect vehicles in the visual domain using audio input. 
Their models were trained with pseudo-labels obtained via pre-trained visual networks: 
Gan \textit{et al.} \cite{Gan2019SelfSupervisedMV} and Rivera Valverde \textit{et al.} \cite{valverde:2021:mmdistillnet} used respectively a stereo microphone and a microphone array to estimate 2D bounding boxes, while Vasudevan \textit{et al.} \cite{vasudevan:2020:semanticobject} used binaural sound for semantic segmentation of 360° street views. 
In contrast, we use a 16-channel microphone array and deal with human speakers, not cars. 
Unlike continuous car noise, speech is challenging as detected faces may be actively speaking or silent.
Thus, active speaker localization combines activity detection with source localization which is a problem of major interest especially to the audio community, see for instance the Task3 of the DCASE challenge \cite{politis:2020:DCASE}.

According to Yann LeCun's definition \cite{LeCun:2019:tweet}, 
we use \textit{self-supervised learning} in that part  of the input (visual frames \& mono audio) supervises a network fed the remainder (multi-channel audio). 
Yet, it is also \textit{semi-supervised} in that a supervised teacher model, pre-trained on a labeled dataset, automatically annotates unlabeled data to train the student network \cite{Zhu:2009:semiSupervised}.

\subsection{Active Speaker Detection}

Active speaker detection (ASD) is typically tackled as a multi-modal learning problem: e.g.,
via correlation between voice activity and lip  
or upper body motion \cite{Cutler:2000:lookWho,Haider:2012:towardsspeaker,Chakravarty2015WhosSA}. 
The first, large, annotated dataset for ASD was introduced for the ActivityNet Challenge (Task B) at CVPR 2019: the AVA-ActiveSpeaker dataset \cite{Roth:2020:AVA}. 
It has 38.5 hours of face tracks and audio extracted from movies, labeled for voice activity. 
Chung \textit{et al.} won the challenge
with an audio-visual model pre-trained on audio-to-video synchronization, performing 3D convolutions \cite{Chung2019NaverAA}. 
In 2020, Alcázar \textit{et al.} proposed Active Speakers in Context (ASC) \cite{Alcazar_2020_CVPR}. Instead of compute-intensive 3D convolutions or large-scale audio-visual pre-training, ASC leverages  context: in assessing a face's speech activity, it looks at any other faces.
While highly-effective, these audio-visual classifiers rely on prior visual face extraction.
In practice, the speaker can be occluded or facing away from the camera, causing face extraction to fail and degrade overall system performance.
We overcome this using multi-channel audio input: at inference, the model relates voice activity to the speaker's position in the visual frame by combining deep learning with beamforming techniques to optimize the localization accuracy. 
ASD can also be tackled with 'traditional' (other than deep learning) solutions. For example, Mohd Izhar \textit{et al.} \cite{izhar:2020:AVtracker} proposed a 3D audio-visual speaker tracker that combines 3D visual detections of the speaker's nose, estimated with the aid of multiple cameras, with 3D audio source detections, estimated using a microphone array and steered response power of the acoustical signal.

Here we present four original contributions: (1) a cross-modal student-teacher approach in which we train the multi-channel audio network with pseudo labels from a pre-trained audio-visual teacher network; (2) spatial feature extraction by beamforming prior to the conventional log-mel-spectrogram input to the audio network;
(3) training and evaluation with an audio-visual dataset captured with multiple co-located cameras and microphone array;
(4) substantial gain in performance by the audio-beamforming student network over all baseline audio networks and the audio-visual teacher network, halving the overall error rate. In light of the excellent results achieved, future benchmark will also compare our solution with other state-of-the-art approaches, e.g. \cite{izhar:2020:AVtracker}.

\section{Self-Supervised Approach} \label{appr}

\begin{figure}[tb]
\centerline{\includegraphics[width=\columnwidth]{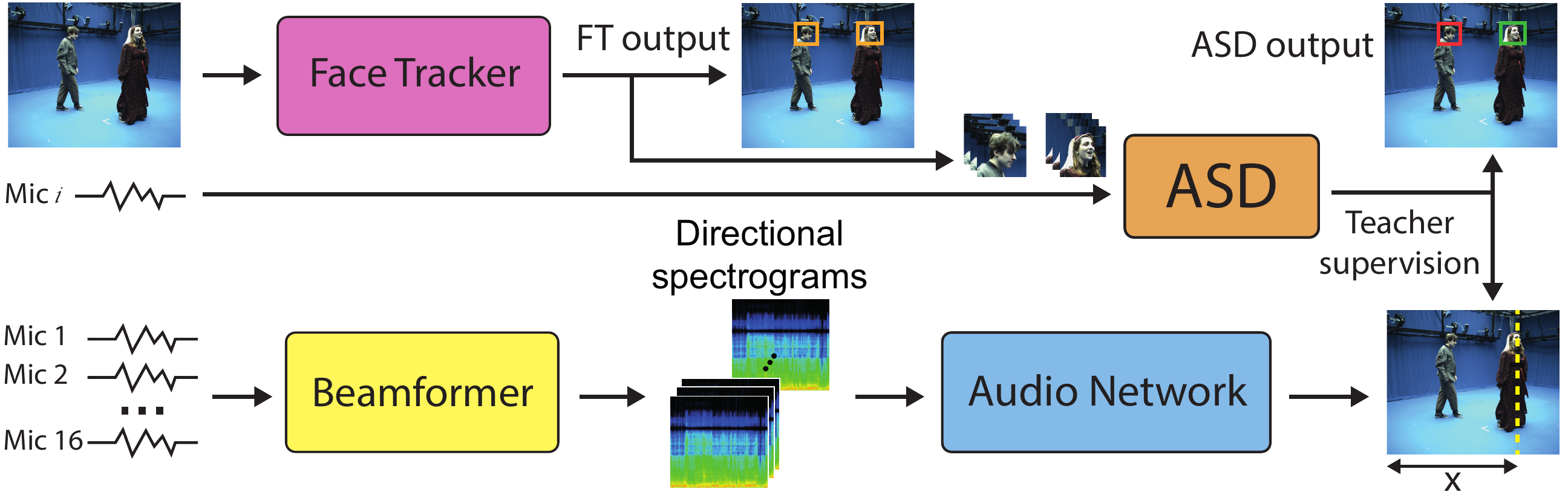}}
\caption{Block diagram of the proposed learning pipeline. The `teacher' network, composed of an audio-visual active speaker detector preceded by a face tracker, produces pseudo labels with the speaker's position at each time instance. The `student' audio network is trained to regress the speaker's horizontal position (x) from the `directional' spectrograms achieved after filtering with the spatial beamformer.}
\label{block_diagram}
\end{figure}

Our machine-learning pipeline consists of an audio-visual teacher and an audio-only student, as in Fig.\,\ref{block_diagram}. 
The teacher network visually detects and tracks faces in the video, then classifies active/inactive frames with the aid of monaural (single-channel) audio.
The horizontal position of the active faces' bounding boxes were used as pseudo labels to train the student network, 
which has multi-channel audio input from a microphone array. 

\begin{figure}[tb]
\centerline{\includegraphics[width=\columnwidth]{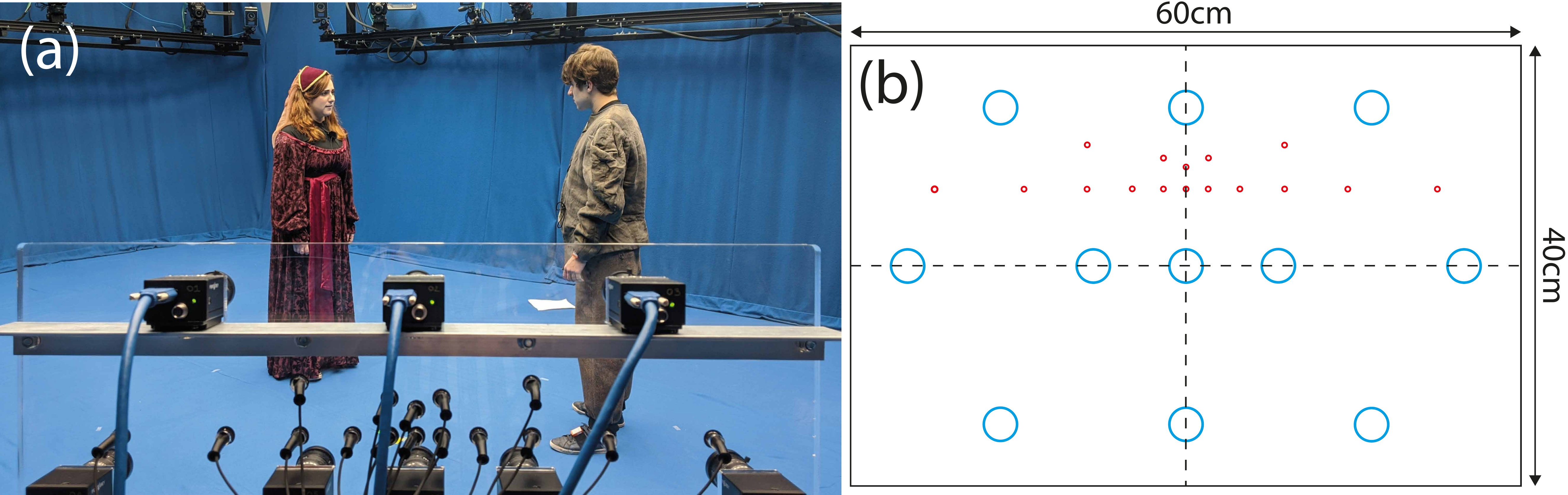}}
\caption{(a) A photo taken from behind one of the two rigs during the dataset capture. (b) Schematic of camera (blue) and microphone (red) positions on the AVA rig.}
\label{AVArig}
\end{figure}

\subsection{Audio-Visual Speaker Dataset}
To provide audio-visual data for training and testing, spoken scenes were recorded in our studio by two twin Audio-Visual Array (AVA) rigs designed for media production. 
Each AVA rig 
has a 16-element microphone array and 11 cameras fixed on a flat perspex sheet,
as in Fig.\,\ref{AVArig}. 
Each camera captures 2448$\times$2048p video frames at 30\,fps. 
Audio is sampled at 48 kHz, 24 bits. 
The microphone array has a horizontal aperture of 450\,mm 
and vertical aperture of only 40\,mm, thus
prioritizing horizontal spatial resolution.
Horizontally, the microphones are log-spaced for broad frequency coverage from 500\,Hz to 8\,kHz. 
Although our pipeline requires a single video feed with the microphone array, capturing multiple views allows us extend the audio network's training via an additional task: choosing from which view we want the speaker to be localized. 
Hence, we append a one-hot vector denoting the selected view to the audio input, which provides data augmentation through variations in the camera perspective.
The captured scenes are excerpts from Shakespeare's `Romeo and Juliet' played by two actors, seen in Fig.\,\ref{AVArig} (a). 
Captured scenarios, 20 sequences in total of 15--40\,s duration, included monologue, dialogue and physical interaction with occlusion.
As each sequence was captured by 22 cameras in all, the dataset provided over 2.5 hours of video data.

\subsection{Teacher Network}
The teacher network has two main parts: a face tracker and an ASD classifier. 
To generate the face tracks (i.e., stack of face crops), the SeetaFaceEngine2 face detector was first applied to each video frame \cite{wu2016Seeta}.
It proved effective in detecting faces in challenging conditions, such as partial occlusion or faces in profile, and fast (it processed 2448$\times$2048p frames at 10$^+$\,fps). 
The per-frame detections were tracked over time based on bounding-box intersection over union (IoU) across adjacent frames.
Gaussian temporal smoothing was applied to the bounding box corners.
For ASD, we trained the publicly-available ASC model \cite{Alcazar_2020_CVPR} on the AVA Active Speaker dataset \cite{Roth:2020:AVA}. 
In \cite{Alcazar_2020_CVPR}, the \textit{i-th} frame is classified observing the previous and following \textit{n} frames, input as a stack of 2$n$+1 frames. 
Experiments on our data performed best with stacks of 5 frames ($n$\,=\,2). 
Once trained, the model processed our dataset to automatically detect the active speakers. 
The horizontal positions of the speakers' bounding-box centers yielded pseudo labels to supervise the training of the student audio network.

\subsection{Student Network}
The student network is trained to tackle active speaker detection as a simultaneous classification and regression problem: active/inactive classification, and regression of the speaker's position when speech activity is detected. 
The network is trained using active and silent segments. 
To localize moving vehicles, Gan \textit{et al.} \cite{Gan2019SelfSupervisedMV} fed their network directly with magnitude mel-spectrograms of the left and right stereo microphone signals, neglecting phase information and emphasizing level-difference cues.  
With microphone arrays, we argue that fine time-difference cues are key for direction-of-arrival (DoA) estimation.
So, our solution spatially filters the array's audio by steering its directivity over \textit{N} horizontal so-called `look' directions. 
The output \textit{N} audio signals give magnitude mel-spectrograms, 
stacked as an \textit{N-}channel image input to the student network. 
The advantages of this solution are twofold: time-difference cues are kept, and the beamformer's steered outputs help the network to converge.

\begin{figure}[tb]
\centerline{\includegraphics[width=\columnwidth]{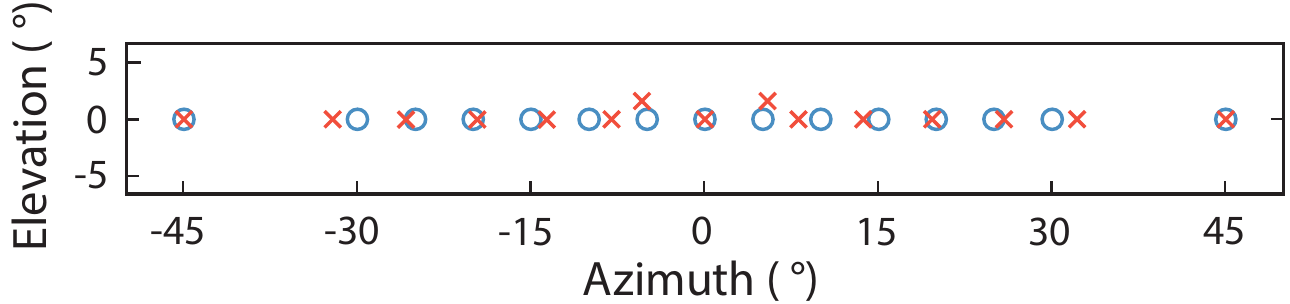}}
\caption{The 15 look directions employed. Blue circles represent the ideal steering angles; Red crosses are the Lebedev sampling grid approximations.}
\label{look_dir}
\end{figure}

\textbf{Beamformer:} Using the beamforming toolbox by Galindo \textit{et al.} \cite{galindo:2020:microphone}, a set of beamforming weights was computed to spatially filter the array signals over 15 look directions. We employed the super-directive beamformer (SDB) with white noise gain constraint as it preserves excellent quality of the target sound with good suppression of external noises \cite{galindo:2020:microphone}.
The designed frontal look directions are equally spaced along the horizon by 5° from one another in the range $\pm$30°, plus two additional directions at $\pm$45°. These positions were approximated by a Lebedev sampling grid, which has nearly uniform distribution over the sphere but is non-uniform in the azimuth-elevation representation seen in Fig.\,\ref{look_dir}.

\begin{figure}[tb]
\centerline{\includegraphics[width=\columnwidth]{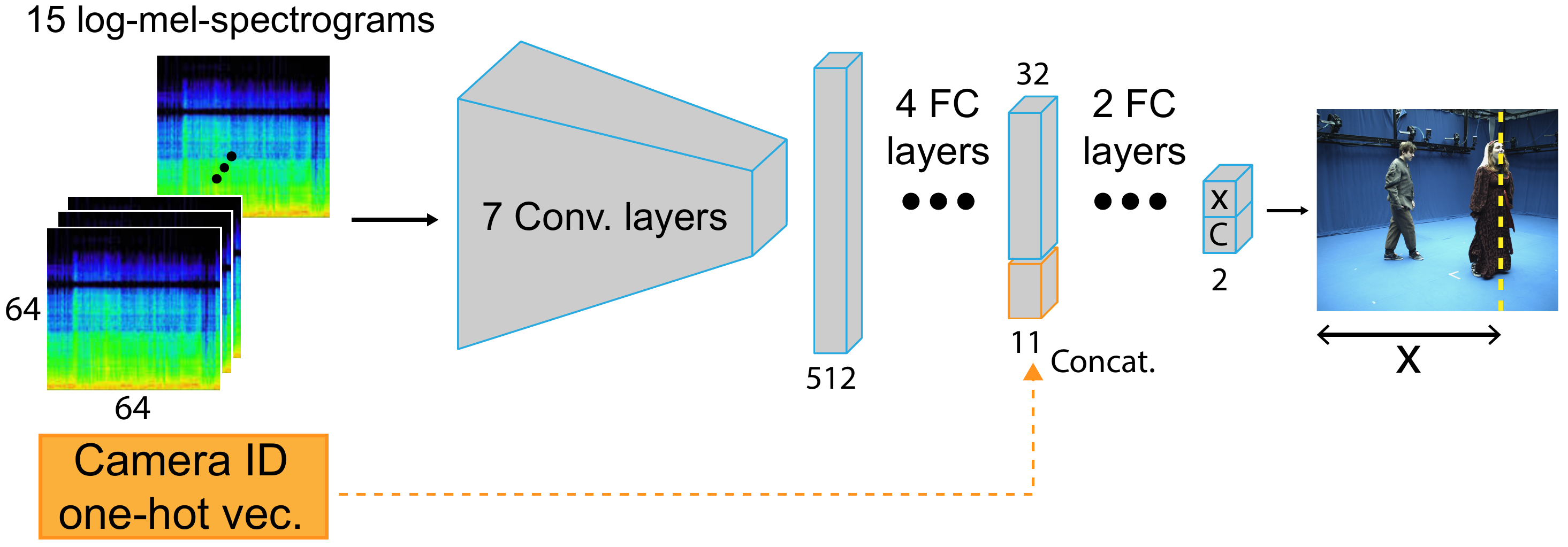}}
\caption{Schematic representation of our `student' audio network.}
\label{audio_network}
\end{figure}

\textbf{Model:} The audio network takes as input a stack of 15 log-mel-spectrograms computed from the 15 beamformer output signals. 
The network is depicted in Fig.\,\ref{audio_network}. 
It has 7 convolutional layers that progressively decrease the time-frequency resolution of the spectrograms and increase the number channels until a 1$\times$1$\times$512 feature vector is achieved. 
Then, the feature vector passes through 4 fully connected layers. 
The resulting vector is concatenated with an 11-dimensional one-hot vector encoding the camera view with which to perform the regression. 
The reason for performing the concatenation at this stage of the network is to first reduce the length of the feature vector to approximate that of the one-hot vector. 
Lastly, 2 fully-connected layers generate a 2D output vector representing the speaker's horizontal position and the confidence. 
Each convolutional layer is followed by a ReLU layer to introduce non-linearity, dropout of 20\%, and batch normalization. 
Similar to \cite{Krizhevsky:2012:AlexNet}, after the first, second, fifth and sixth layers, max pooling layers with stride 2 are applied to rapidly decrease the feature map dimensionality. 
The output coordinates and confidence are normalized in the range [0, 1] by means of a Sigmoid function.
The per-frame detections are then temporally filtered using Gaussian smoothing to provide temporal coherence.

\textbf{Loss Function} To train our model, we employ a sum-squared error-based loss function. 
The loss is composed of two terms, a regression loss and a confidence loss:
\begin{equation}
Loss=\mathbbm{1}\textsubscript{src}(x-\hat{x})^2+(C-\hat{C})^2
\label{loss_function}
\end{equation}
where \textit{\^{x}} and \textit{x} are respectively the predicted and ground truth positions of the speaker along the horizontal axis of the video frame, normalized in the range [0, 1], while \textit{\^{C}} and \textit{C} are the predicted and ground truth confidences. 
While the ground truth position of the speaker is simply provided by the teacher network, the confidence is computed as follows:
\begin{equation}
  C =
    \begin{cases}
      1-|(x-\hat{x})| & \text{if source is active}\\
      0 & \text{otherwise}\\
    \end{cases}       
\end{equation}
In doing so, the confidence is forced to be low in the absence of speech activity, and to estimate how close the coordinate prediction is to the source. 
The farther the coordinate prediction from the source, the lower the confidence is forced to be. 
This means that \textit{\^{C}} represents the confidence that a source is in the model's predicted position. 
The term $\mathds{1}$\textsubscript{src} is 0 for silent frames and 1 for active, i.e., in the absence of speech the network is penalized by the confidence loss only.

\section{Experiments} \label{exp}

\subsection{Data Preparation} 
To train the audio network we use short segments of audio (audio frames) temporally centered on the visual time frame of interest.
To increase the reliability of the positive face tracks detected by the teacher network, we manually screened the positive pseudo labels in order to remove potential false positives. In such a way, we ensure that all the active frames used in training are correct. However, we can not ensure that the negatives are actually true, i.e. that they were classified as silent because of the effective absence of speech or simply because the speaker's face was not detected. Therefore, we used the AVA rigs to capture a silent audio sequence containing only the studio noise floor. The audio frames employed for training are sampled either from the teacher network's active detections or from the silent sequence to ensure the training samples are reliable. The silent sequence was captured to be long enough to approximately match the number of speech frames and therefore avoid bias between activity and inactivity.      
In total we collated a training set with over 140k frames from 17 of the 20 dataset's sequences and silent segments. 
The remaining 3 unseen sequences were used only at inference, for testing. To perform an impartial evaluation, we manually labeled the ground-truth bounding-box locations of the speaker's head in each of the 22 camera views, to yield 66 test sequences.

\subsection{Implementation Details}
As in \cite{vasudevan:2020:semanticobject}, gains were computed for each channel of the array to set their the root mean square (RMS) magnitude to a desired value and achieve level calibration.
The audio frames used to train the network are 5 visual frames long, i.e., 166\,ms. Therefore, likewise it was done for the training of the teacher network, to evaluate the \textit{i-th} frame, we consider an audio chunk that starts from the frame $i-2$ and finishes at frame $i+2$, (8000 audio samples). We extract the audio chunk of interest from each of the 15 signals achieved with the beamformer. A short-time Fourier transform with a Hann window of size 512 samples is applied at hop steps of 125 samples to generate a spectrogram for each look direction, discretizing the 8000-sample audio chunks into 64 temporal bins. The frequency resolution of the spectrograms is down-sampled over 64 mel-frequency bins and finally, the logarithm operation is applied. Stacking together the 15 log-mel-spectrograms, the result is a 64$\times$64 image with 15 channels. To leverage and amplify the differences between silent and active frames at each frequency band, we normalize the network's inputs following a frequency-wise normalization criterion. In the first place, the mean of the pixels' values is computed for each frequency bin over the entire training set. Then, each pixel value in normalized by subtracting its frequency bin's mean, and dividing by the global standard deviation. We found that this normalization approach enhances the performance in comparison to the traditional normalization approach based on the global mean and global standard deviation. 
We trained our audio network for 25 epochs using batches of 64 images, Adam optimizer, and a learning rate of \num{2e-4}.

\subsection{Evaluation Metrics}
A prediction is considered to be positive, i.e. the network predicts the presence of speech, when the predicted confidence is above a threshold and a positive detection is considered to be true when the localization error is within a predefined tolerance. 
We compute the precision and recall rates by varying the confidence threshold from 0\% to 100\%. Since, at inference, the network's confidence tends to be either very high or very low, we build the precision-recall curves by sampling the thresholds from a Sigmoid-spaced distribution. This provides more data points for high and low confidence values. The common object detection metric average precision (AP) was then computed. The AP is determined following the approach indicated by the Pascal VOC Challenge \cite{Everingham:2015:pacalVOC}, which consists in: 1) compute a monotonically decreasing version of the precision-recall curve by setting the precision for the recall \textit{r} equal to the maximum precision obtained for any recall \textit{r'$\geq$r}, and 2) compute the AP as the numerical integration of the curve, i.e. the area under the curve (AUC). 

According to human auditory spatial perception \cite{Strybel:2000:MinimumAA}, the minimum audible angle (MAA) is 2°. Therefore, we set a tolerance for spatial misalignment between the prediction and the GT speaker's position of $\pm$2° along the azimuth, which corresponds to $\pm$89 pixels if projected onto the image plane\footnote{All conversions from pixels to degrees and vice-versa are performed with pre-computed camera-calibration data.}. 
However, in many practical audio-visual applications, the human brain tends to tolerate greater misalignments as it integrates audio and visual signals as a unified object (ventriloquism effect \cite{Stenzel:2018:PTC,Berghi:2020:IEEEVR}). 
For this reason, the AP is also reported at $\pm$5° tolerance ($\pm$222 pixels) to provide greater flexibility. 
The F1 score to find the optimal compromise between precision and recall is computed too. Additionally, the average distance (aD) between the active detections and the GT locations is reported in both pixel and angle units.
Finally, regardless of the regression accuracy, we measure the amount of correct classifications between active and silent frames achieved by setting the confidence at 0.5. 

\begin{table}[tb]
\caption{Method results in average distance (aD),  average precision (AP) and F1 score at 89 ($\pm$2°) and 222 ($\pm$5°) pixel tolerances.}
\begin{center}
\begin{tabular}{c|c|c|c|c|c}
\hline
\textbf{Method}&\textbf{aD}&\textbf{AP@2°}&\textbf{F1@2°}&\textbf{AP@5°}&\textbf{F1@5°} \\
\hline
Teacher &\textbf{\hl{10p {(0.2°)}}} & 61.1\% & 0.68 & 61.2\% & 0.68 \\
Mono & \hl{318p {(7.2°)}} & 5.1\% & 0.19 & 23.0\% & 0.44\\
Stereo & \hl{196p {(4.4°)}} & 19.3\% & 0.39 & 52.6\% & 0.67\\
w/out BF & \hl{145p {(3.3°)}} & 36.2\% & 0.54 & 71.5\% & 0.79\\
3 look dir & \hl{86p {(1.9°)}} & 53.9\% & 0.69 & 82.7\% & 0.88\\
7 look dir & \hl{77p {(1.7°)}} & 59.0\% & 0.72 & 86.6\% & 0.90\\
Ours & \hl{76p {(1.7°)}} & 66.6\% & 0.79 & 86.2\% & 0.91\\
Ours + tc & \hl{68p {(1.5°)}} & \textbf{70.5\%} & \textbf{0.81} & \textbf{87.1\%} & \textbf{0.92}\\  \hline 
\end{tabular}\vspace{-3ex}
\label{tabBaselines}
\end{center}
\end{table}

\subsection{Baseline Methods} \label{baseln}
We evaluate our approach performance against a variety of different baselines to motivate its design choices.

\textbf{Teacher network:} We process the test set sequences with the teacher network. Note that while our approach estimates only the horizontal position of the speaker, the teacher provides the bounding box coordinates of the speaker's face. Therefore, for a fair comparison, we consider only the x coordinate of the center of the predicted bounding box. 

\textbf{W/out BF:} To prove the benefit introduced by the use of the beamformer, we train a model to perform the same task using directly the log-mel-spectrograms of the 16 microphones signals. The model used is identical but the input images have 16 channels rather that 15. 

\textbf{Stereo:} Similarly to the `W/out BF' baseline, we do not apply spatial filtering to the audio signal. In this case, we use the log-mel-spectrograms achieved from just 2 of the microphones. We use the microphones spaced at $\pm$88.3 mm distance from the center of the array, which is consistent with the ORTF stereo microphones capturing technique. In this case, the inputs of the network are 2-channels images.

\textbf{Mono:} We only use the log-mel-spectrogram of the central microphone signal. This allows to observe what the model can learn from monaural sound and no additional spatial cues.

\textbf{3 \& 7 look dir:} The audio network is trained with a fewer number of look directions to observe the effects of a coarser input spatial resolution. The `3 look dir' baseline employs only the look directions at 0° and $\pm$20°; while for the `7 look dir' the directions at 0°, $\pm$15°, $\pm$30° and $\pm$45° are used. 

\textbf{Ours + tc:} We smooth the per-frame consecutive detections achieved by means of a Gaussian temporal filter that provides temporal coherence.

\section{Results} \label{res}

\subsection{Method Comparisons}
\begin{figure}[tb]
\centerline{\includegraphics[width=\columnwidth]{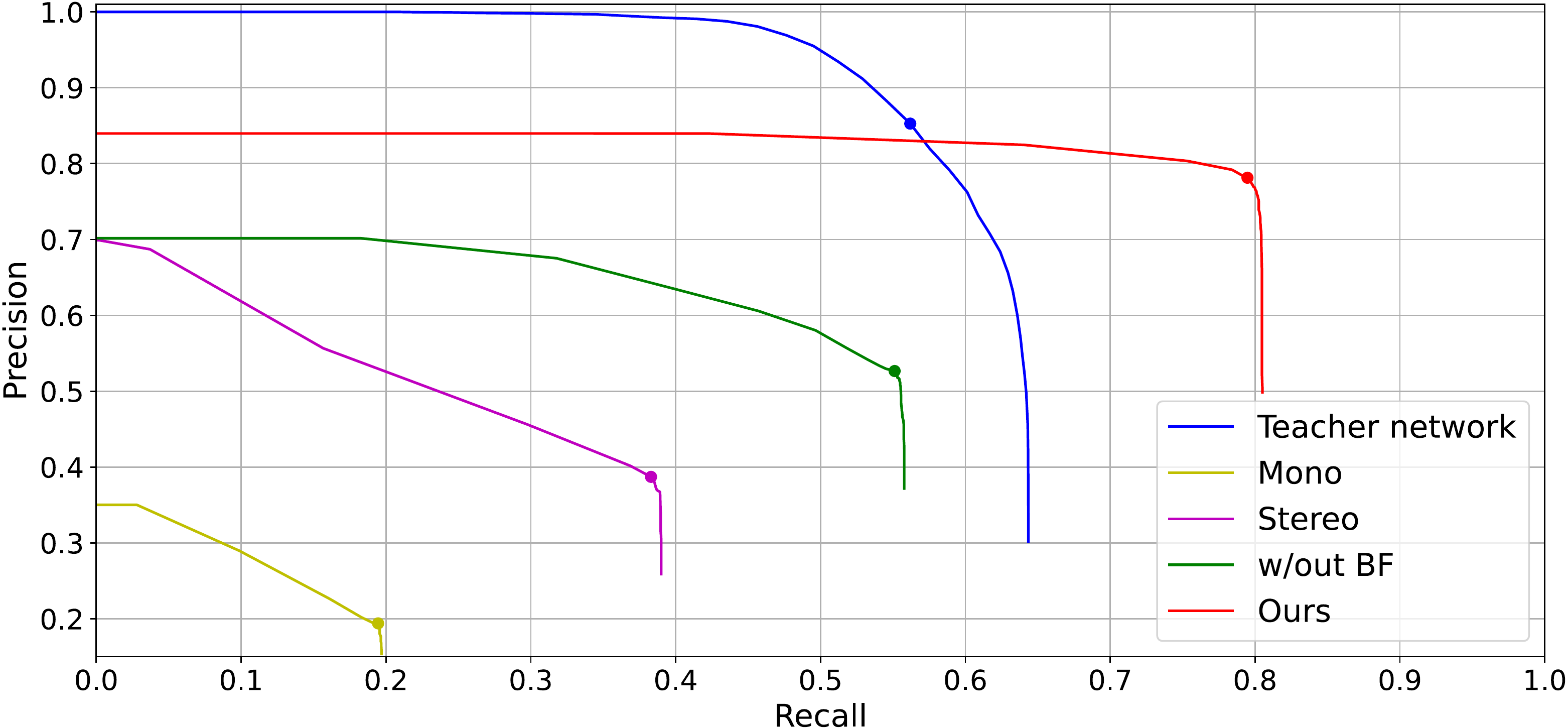}}
\caption{Method comparison of precision versus recall at 89-pixel tolerance. Dots mark the highest F1 scores.}
\label{precision_recall}
\end{figure}

Experimental results show that our approach considerably outperforms the other baselines. In Fig.~\ref{precision_recall} are reported the precision-recall curves achieved with the different baselines, while the metrics results are reported in Tab.~\ref{tabBaselines}. 

\textbf{Ours vs. Teacher network:} As appreciable from Fig.~\ref{precision_recall}, our approach produces a remarkable improvement in recall rate, meaning that it detects the speakers more easily compared to the teacher network. The teacher network's poorer recall rate is caused by its impossibility to detect speakers when their face is not visible to the face detector, e.g., when the actors are not facing the camera. This problem does not persist with our audio network. Contrariwise, the spatial resolution achievable by the visual modality produces a higher maximal precision, although overall our approach has greater AP.   

\begin{figure}[tb]
\centerline{\includegraphics[width=\columnwidth]{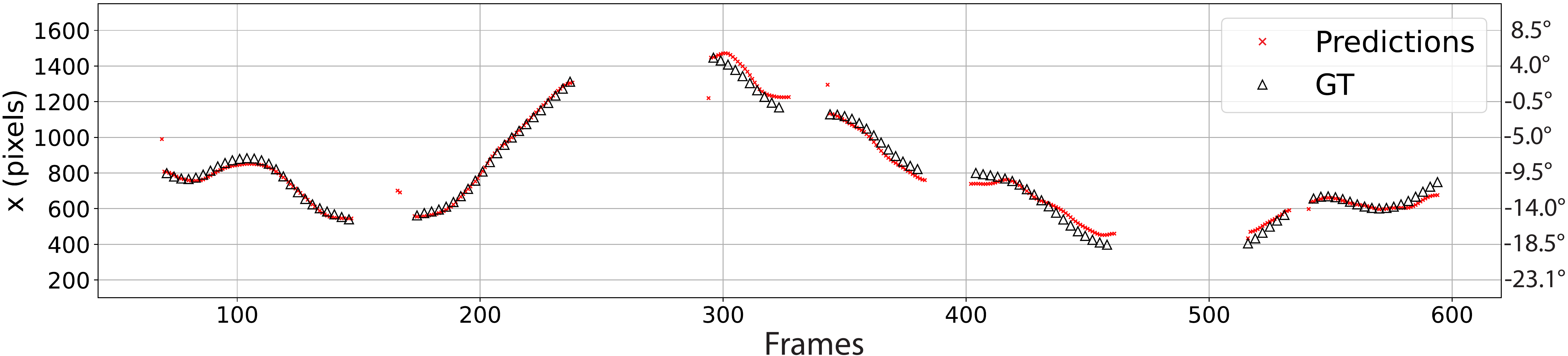}}
\caption{Example of detections over time from one camera view on a test sequence. Pixel coordinates provided also in angles for reference.}
\label{azimuthOverTimeMonologue}
\end{figure}

\textbf{W/ vs. W/out Beamformer:} The employment of the beamformer introduces an important improvement over the direct use of the audio channels, with an increment in AP of 30.4 percentage points at 89p tolerance and 14.7 percentage points at 222p. Benefits at both the precision and recall rates are noticeable and the average distance is decreased by \hl{69p ({1.6°} shorter)}. Even with fewer look directions the benefit is remarkable, meaning that the beamformer provides the overall system with a higher localization accuracy. The advantage of additional look directions is less appreciable with a larger tolerance (222p) where the AP resulted slightly higher with 7 look directions than with 15.

\textbf{16 Channels vs. Stereo vs. Mono:} The availability of a higher number of microphones improves the network's spatial resolution in that it provides additional spatial cues. The employment of 16 microphones rather than only 2 decreases the aD error from \hl{196p {(4.4°)} to 145p {(3.3°)}} and, as a result, the precision and recall rates are higher. Nevertheless, the Stereo configuration still proved to deliver a decent degree of spatial accuracy which might be useful in situations with limited availability of microphones. As expected, the Mono baseline produced a poor AP, although its aD error might appear to be relatively low. This is caused by the network trying to minimize the regression error by predicting the speaker's position to always be towards the center of the visual frames. 

\subsection{Spatial Resolution}
The results show that the spatial resolution increases when more look directions are employed, however, the improvement is not remarkable. On average, \hl{10p} improvement from 3 to 15 look directions, and only 1p improvement from 7 to 15 look directions.
The spatial resolution can be further refined by temporally filtering the per-frame detections in order to provide temporal coherence as defined in \ref{baseln}. After the filtering, the average distance error is only \hl{68p {(1.5°)}}. 
Also, it would be hard to keep the error below this value considering that a certain degree of approximation is already introduced by the spatial offset between the mouth and center of the bounding box itself. As expected, the spatial resolution achievable with the teacher network is very high thanks to the visual modality.

\subsection{Speech Activity Detection}
In correctly classifying the presence of speech, all the baselines are approximately equally highly performant: the `w/out BF' baseline classified 94,7\% of the frames correctly, and both our approach and Stereo 95.7\%.  
In fact, to accomplish such a task the use of beamformers and multiple audio channels results in being superfluous and potentially redundant. Indeed the highest classification performance was achieved by the Mono baseline with 96.6\% correctness. By observing the false positives detections, it is possible to notice that they are mainly caused by external noises such as foot steps and actors' breath in prior to the speech segment, meaning that the network learned to distinguish between the presence of general activity and silence, while the manual annotations only reported the actual speech segments. This behavior can be explained by the fact that the silence frames used in training came from a completely silent capture of the room noise floor. If a more selective model is needed, it can be accomplished by including unwanted non-speech sounds in the silent track. 

\section{Conclusion} \label{concl}
In this work, we proposed a self-supervised learning approach to solve the active speaker detection and localization problem using a microphone array. To generate the pseudo labels used to train our model from newly collected data, we adopted a pre-trained audio-visual teacher network. We proposed a new strategy for combining spatial filtering with deep learning techniques and evaluated our system against several different baselines, with a favorable outcome. Future experiments will explore different beamformers, look directions, and network architectures, as well as benchmark our approach with other state-of-the-art solutions and datasets.

\section*{Acknowledgment}

Thanks to Marco Volino, Mohd Azri Mohd Izhar, Hansung Kim, Charles Malleson and actors for audio-visual recordings.

\bibliographystyle{IEEEtran}
\bibliography{IEEEMMSP2021}



\end{document}